\documentclass{aa}  

\usepackage{graphicx}
\usepackage{txfonts}
\begin{document} 

   \title{Model comparison for the density structure along\\ solar prominence threads}
   \titlerunning{Density structure along solar prominence threads}
\authorrunning{Arregui \& Soler}
   \author{I. Arregui
          \inst{1,2}
          \and
          R. Soler\inst{3}
          }
   \institute{Instituto de Astrof\'{\i}sica de Canarias, E-38205 La Laguna, Tenerife, Spain
            \and
            Departamento de Astrof\'{i}sica, Universidad de La Laguna, E-38206 La Laguna, Tenerife, Spain\\
            \email{iarregui@iac.es}
            \and
            Departament de F\'{\i}sica, Universitat de les Illes Balears, E-07122 Palma de Mallorca, Spain\\
            \email{roberto.soler@uib.es}
             }
             
   \date{Received; accepted}


  \abstract
   {Quiescent solar prominence fine structures are typically modelled as density enhancements, called threads, which occupy a fraction of a longer magnetic flux tube. This is justified from the spatial distribution of the imaged plasma emission/absorption of prominences at small spatial scales.  The profile of the mass density along the magnetic field is however unknown and several arbitrary alternatives are employed in prominence wave studies. The identification and measurement of period ratios from multiple harmonics in standing transverse thread oscillations offer a remote diagnostic method to probe the density variation of these structures.}
   {We present a comparison of theoretical models for the field-aligned density along prominence fine structures. They aim to imitate density distributions in which the plasma is more or less concentrated around the centre of the magnetic flux tube.  We consider Lorentzian, Gaussian, and parabolic profiles. We compare their theoretical predictions for the period ratio between the fundamental transverse kink mode and the first overtone to obtain estimates for the ratio of densities between the central part of the tube and its foot-points and to assess which one would better explain observed period ratio data.}
   {Bayesian parameter inference and model comparison techniques are developed and applied. Parameter inference requires the computation of the posterior distribution for the density gradient parameter conditional on the observable period ratio. Model comparison involves the computation of the marginal likelihood as a function of the period ratio to obtain the plausibility of each density model as a function of the observable and the computation of Bayes Factors to quantify the relative evidence for each  model, given a period ratio observation.}
   {A Lorentzian density profile, with plasma density concentrated around the centre of the tube seems to offer the most plausible inversion result. A Gaussian profile would require unrealistically large values of the density gradient parameter and a parabolic density distribution does not enable us to obtain well constrained posterior probability distributions. However, our model comparison results indicate that the evidence points to the Gaussian and parabolic profiles for period ratios in between 2 and 3, while the Lorentzian profile is preferred for larger period ratio values.  The method here presented can be beneficial to obtain information on the plasma structure along threads, provided period ratio measurements become widely available.}
{}


   \maketitle

\section{Introduction}
When observed with high spatial resolution instrumentation, the fine structure of solar prominences appears to be distributed in the form of fine strands of plasma presumably aligned with the local magnetic field. This strands become visible as bright/dark filamentary structures in emission/absorption when observing prominences in the limb/on the disk.  As described by \cite{heinzel07}, the fine structure of prominences is manifested differently in the case of limb observations and on disk filaments thus making difficult to identify the same structures as seen on the limb and on the disk. When observed on the limb, the fine structure is composed by horizontal and vertical threads and knots of prominence plasma that exhibit a significant dynamical behaviour \citep[see e.g.,][]{heinzel07,berger08,mein91,schmieder10,schmieder13,schmieder14}. On the disk, the high-resolution $H_\alpha$ images show fine-structure fibrils, visible along the spine of  quiescent filaments, with average widths of the order of 0.3\arcsec, whereas their length is in between 5\arcsec and 40\arcsec \citep{lin05,lin11}. Filament threads are thought to be made of relatively cold and dense plasma, with chromospheric properties, occupying a fraction of a longer invisible magnetic flux tube and surrounded by hot plasma with coronal properties. 

In the last years, the analysis of spectra and images of prominence plasmas and the improvement of non-LTE radiative transfer models have enabled the development of inversion and forward modelling techniques to infer plasma parameters in solar prominences \citep[see][for a review]{labrosse10}. \cite{tandberg-hanssen95} and \cite{patsourakos02} give a compilation of various density determinations in prominences and show that this parameter varies by at least 2 orders of magnitude, from 10$^{9}$ to 10$^{11}$ cm$^{-3}$. Magnetic fields show a similar variability with field strengths that range from a few G to 20-30 G in quiescent prominences \citep[e.g.,][]{bommier94,merenda06,gunar07} to higher values in active region prominences. Spectro-polarimetric observations are now able to offer information on the prominence magnetic field, although individual flux tubes cannot yet be resolved  \citep{lopezariste02,lopezariste05,orozco14}. These studies have provided a wealth of knowledge on the physical quantities and dynamics of prominences \citep{heinzel06,heinzel07,mackay10,labrosse10,parenti14,vial15}.

To obtain information about the value of physical quantities,  such as the temperature, density, and magnetic field strength, and their spatial variation along and across the magnetic field is of fundamental importance to understand the physical properties and dynamical processes operating in solar prominences. A widely followed approach is the 1D and 2D non-LTE modelling of prominences and filaments, including their fine structures, and the comparison between the observed spectral line properties with the synthetic line characteristics \citep[e.g.,][]{heinzel95,paletou95,gouttebroze00,gunar07,gunar13,gunar14,heinzel15}. An alternative approach is the theoretical modelling of magnetohydrodynamic (MHD) waves in fine-structure equilibrium models and the comparison of the observed waves with the theoretical MHD wave properties, a diagnostic tool known as prominence seismology \citep{ballester05,ballester06,oliver09,ballester14}.

The method of prominence seismology relies on the comparison between observed and theoretical properties of waves and oscillations. The analysis of imaging and spectroscopic data has clearly shown the presence of small amplitude oscillations in prominence fine structures \citep[see][for a review]{arregui12a}. Some of these oscillations produce the transverse displacement of the threads and have been  interpreted in terms of MHD transverse kink waves \citep{lin09,lin11}. The observed disturbances have characteristic periods of a few minutes and velocity amplitudes of a few km s$^{-1}$. By measuring the oscillation properties and confronting them to theoretical predictions information on the unknown parameters can be obtained. Because of the relatively simple structure of prominence threads, in comparison to the full prominence structure,  seismology studies approximate their magnetic and plasma configuration using simplified magnetic flux tube models in cylindrical geometry. By adopting the zero plasma-$\beta$ approximation, models that study linear tube oscillations are then constructed by specifying a particular density structure that defines the thread inside the tube. Following this approach, models have been used to analyse the oscillatory properties of standing transverse thread oscillations. \cite{diaz02} use a piece-wise constant density profile along and across the thread to model a cold thread embedded in a hotter coronal gas. \cite{soler10b} extended this model by considering a smooth variation of density across the field, in order to study damped transverse oscillations and \cite{arregui11c}  further increased the complexity of the modelling by adopting fully non-uniform density models with a smooth variation of density along and across the magnetic field. In all these studies, the density structuring along and across the field was arbitrarily prescribed.

In a recent study,  \cite{soler15} have considered three alternative models for the density profile along threads and have computed period ratios between the fundamental kink mode and the first overtone of transverse thread oscillations. Period ratios are sensitive to the details of the density variation along the magnetic field, hence the inversion of observed period ratios using longitudinally non-uniform density models provides information on the density profile of prominence threads  \cite[see e.g.,][]{diaz10}. Following this approach, and using their theoretical predictions, \cite{soler15} obtain estimates of the magnitude of the density gradient along the thread.

Because any inference is model dependent, their results are conditional on the specific models that have been assumed to explain observations. A second level of inference consist in presenting the same data to different models to assess in a quantitative manner which one is favoured. This paper presents the solution to that problem, which necessarily demands the Bayesian solution to the model comparison problem.  The Bayesian formalism for inference and model comparison is the only fully correct way we have to obtain information about physical parameters and the plausibility of hypotheses from observations under incomplete and uncertain information \citep[see e.g.][]{trotta08,vontoussaint11}. The three density models adopted by \cite{soler15} are considered to devise a method by which their relative plausibility in explaining period ratio data for transverse thread oscillations can be obtained, hence inferring the degree of evidence for each one.  The method thus developed can be applied to the analysis of future period ratio measurements in prominence threads.

The layout of the paper is as follows. Section~\ref{models} describes the considered prominence thread density models. In Section~\ref{forward} the solutions to the forward problem for transverse thread oscillations are discussed. Seismology results for the inference of the density gradient along threads are shown in Section~\ref{inverse}. In Section~\ref{comparison}, the three models are compared to evaluate which one would better explain observed period ratios.
Section~\ref{conclusions} presents our conclusions.
  
\section{Thread density models}\label{models}
Because of the apparent filamentary structure of prominence plasmas at small spatial scales, individual threads are commonly modelled by means of magnetic flux tube models in which a density enhancement, surrounded by plasma in coronal conditions, occupies a fraction of a longer magnetic flux tube. Although prominences are formed by many threads, observations often show that individual threads oscillate independently.  We neglect the interaction between neighbouring structures 
and consider a model for an individual and isolated structure. Following \cite{soler15}, we consider such a prominence thread model consisting of a straight cylindrically symmetric magnetic flux tube of radius $R$ and length $L$, with the magnetic field directed along the longitudinal coordinate, $z$ \citep[see Figure 1 in][]{soler15}. The ends of the tube are located at $z=\pm L/2$. The magnetic field strength, $B_0$, is uniform and under the plasma-$\beta=0$ approximation the spatial distribution of the mass density, $\rho_0$, can be chosen arbitrarily. This fact is used to consider three alternative density distributions along the equilibrium magnetic field that are to be compared in our study.

Prominence threads are denser near the tube centre than at the foot-points. By denoting the internal density at the centre as $\rho_{\rm i}(0)=\rho_{{\rm i},0}$ and that at the foot-point as $\rho_{\rm i}(L/2)=\rho_{{\rm i},L/2}$, the ratio of densities is given by $\chi=\rho_{{\rm i},0}/\rho_{{\rm i},L/2}$ with $\chi\geq 1$. For $\chi=1$, the thread in homogeneous. The density variation is stronger for larger values of $\chi$.

\begin{figure}
\centering
   \includegraphics[width=0.5\textwidth]{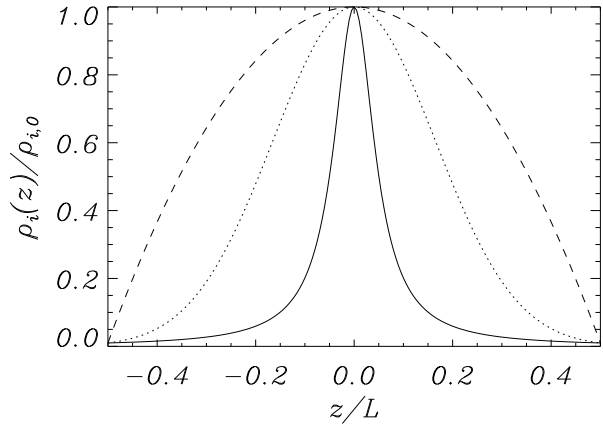}
        \caption{Spatial variation of the equilibrium mass density along the magnetic flux tube for prominence thread models with a Lorentzian profile (solid line), a Gaussian profile (dotted line), and a parabolic profile (dashed line). The ratio of centre to foot-points densities is $\chi=100$.}
              \label{figmodels}
    \end{figure}
 
Using the above definitions, \cite{soler15} construct three alternative density models for longitudinally inhomogeneous prominence threads:  a Lorentzian profile with

\begin{equation}
\rho^{\rm L}_{\rm i}(z)=\frac{\rho_{\rm i,0}}{1+4(\chi-1)z^2/L^2},
\end{equation}
a Gaussian profile with

\begin{equation}
\rho^{\rm G}_{\rm i}(z)=\rho_{\rm i,0}\exp\left(-4\frac{z^2}{L^2} \ln \chi\right),
\end{equation}
and a parabolic profile with

\begin{equation}
\rho^{\rm P}_{\rm i}(z)=\rho_{\rm i,0}\left(1-4\frac{\chi-1}{\chi}\frac{z^2}{L^2}\right).
\end{equation}
Figure~\ref{figmodels} shows the spatial variation of the density distribution for the three considered profiles. For the Lorentzian profile, the dense plasma is located in a narrow fraction of the tube around its centre and the density decreases significantly as soon as we move towards the foot-points of the tube. For the parabolic profile, the density is broadly distributed along the tube. These profiles represent two extreme cases. The intermediate case is given by the Gaussian profile for which the dense plasma is also located around the centre of the tube, but the gradient is smoother that in the Lorentzian case and sharper than in the parabolic case. 

None of the specifically assumed density profiles is expected to be an accurate quantitative representation of the real density variation along solar prominence threads, but they provide a means to model alternative cases which, by means of the model comparison technique presented in this paper, can offer information about whether the density is more or less distributed around the denser central part of the magnetic flux tube.

\section{The period ratio of inhomogeneous threads}\label{forward}
Each of the assumed equilibrium density profiles produces distinct transverse wave signatures. In particular, the ratio of periods between the fundamental transverse kink mode and its first overtone is known to be sensitive to the longitudinal variation of the density along the magnetic flux tube. This is known since the study by \cite{andries05a} in the context of coronal loop oscillations. \cite{andries05b} were the first to use this fact to perform a seismic inversion of the longitudinal density structuring from observations of period ratios of transverse loop oscillations, an approach that has been frequently employed since then \citep[see][for a review]{andries09b}. Period ratio measurements are scarce in prominence oscillations, but  \cite{diaz10} employed the same technique to investigate the dependence of the period ratio on the equilibrium parameters to find that for prominence plasmas the period ratio is larger than two,  and to obtain an estimate of the length of the supporting magnetic tube.

For the particular three density profiles described above, \cite{soler15} obtained analytical and numerical approximations to the period ratio of transverse oscillations of individual threads. The obtained period ratios are functions of a single parameter, the ratio of the central density to the foot-point density.  \cite{soler15} compute period ratios as a function of the increasing density gradient parameter $\chi$ by reducing the foot-point density, while keeping the density at the central part of the tube and find that the period ratio $r=P_0/P_1$ is larger than 2.  They note however that by doing so the total mass in the thread is different for different profiles, making a direct comparison between the results from different profiles unsuitable. To solve this problem, \cite{soler15} consider the same average density in the thread, by defining an average internal density, which keeps the period ratio variation as a function of the average density equal for the three profiles.  The relevant expression is Eq.~(17) in \cite{soler15} which is empirically obtained after analysing their numerical computations. The important result behind this equation is that the period ratio is only a function of the ratio of the average density to the central density. This property enables to express the period ratio as a function of $\chi$, although the functional dependence on this parameter is different for each considered density profile. The relevant forward solutions for each adopted density profile are obtained upon substitution of their expressions (18), (19), and (20) for the average density as a function of $\chi$ into their Equation (17) for the period ratio as a function of the average density \citep[see][]{soler15}. These expressions are

\begin{figure}
\centering
   \includegraphics[width=0.5\textwidth]{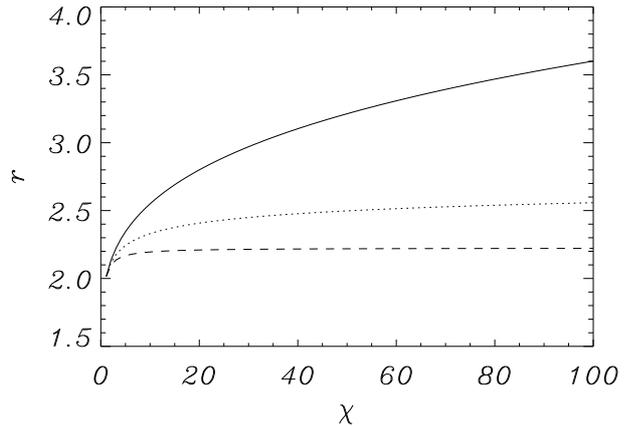}
        \caption{Dependence of the period ratio between the fundamental mode and the first overtone as a function of the density gradient parameter, $\chi$, for a Lorentzian profile (solid line), a Gaussian profile (dotted line), and a parabolic profile (dashed line).}
              \label{figforward}
    \end{figure}

\begin{figure*}
\centering
   \includegraphics[width=0.49\textwidth]{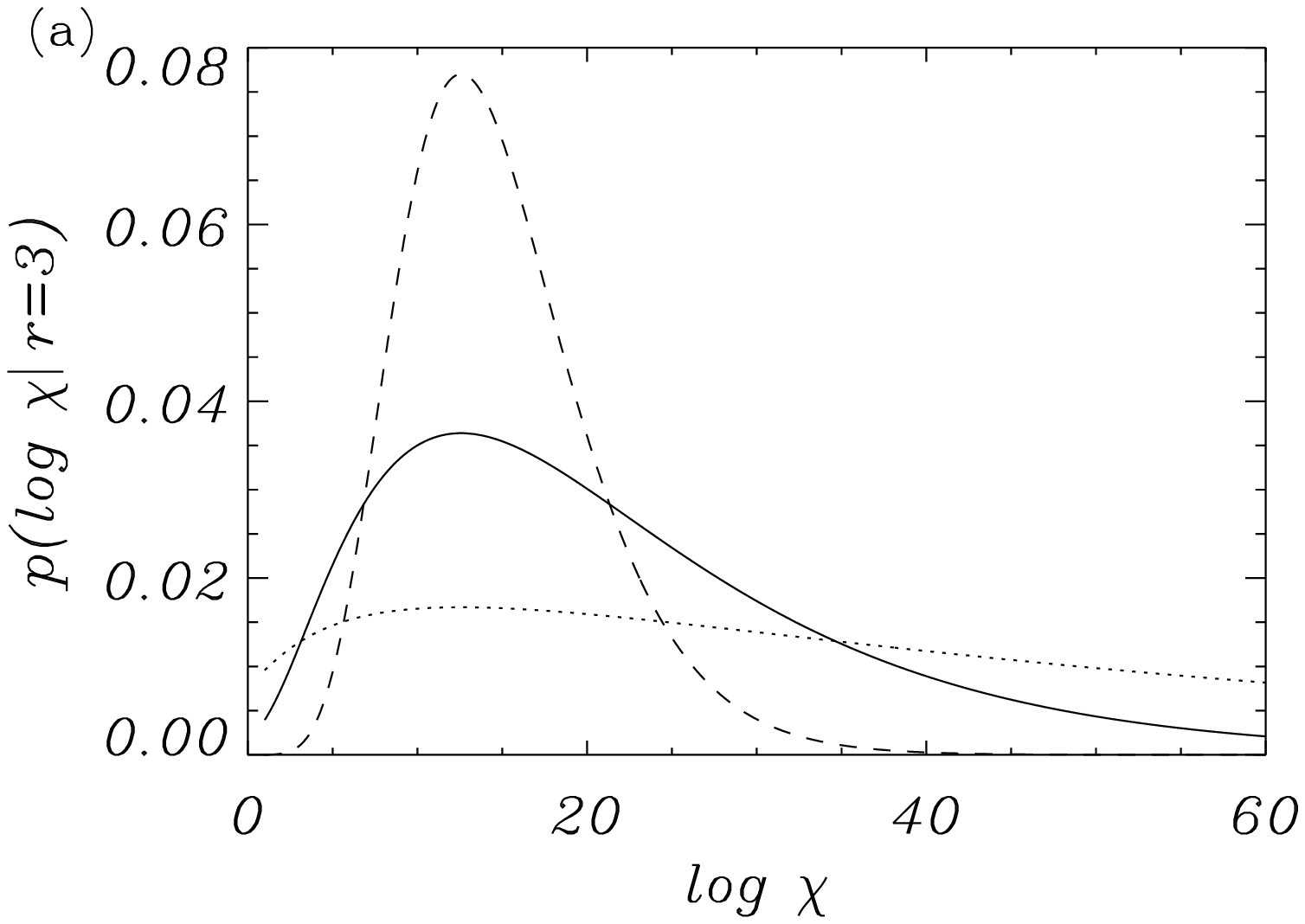}
      \includegraphics[width=0.49\textwidth]{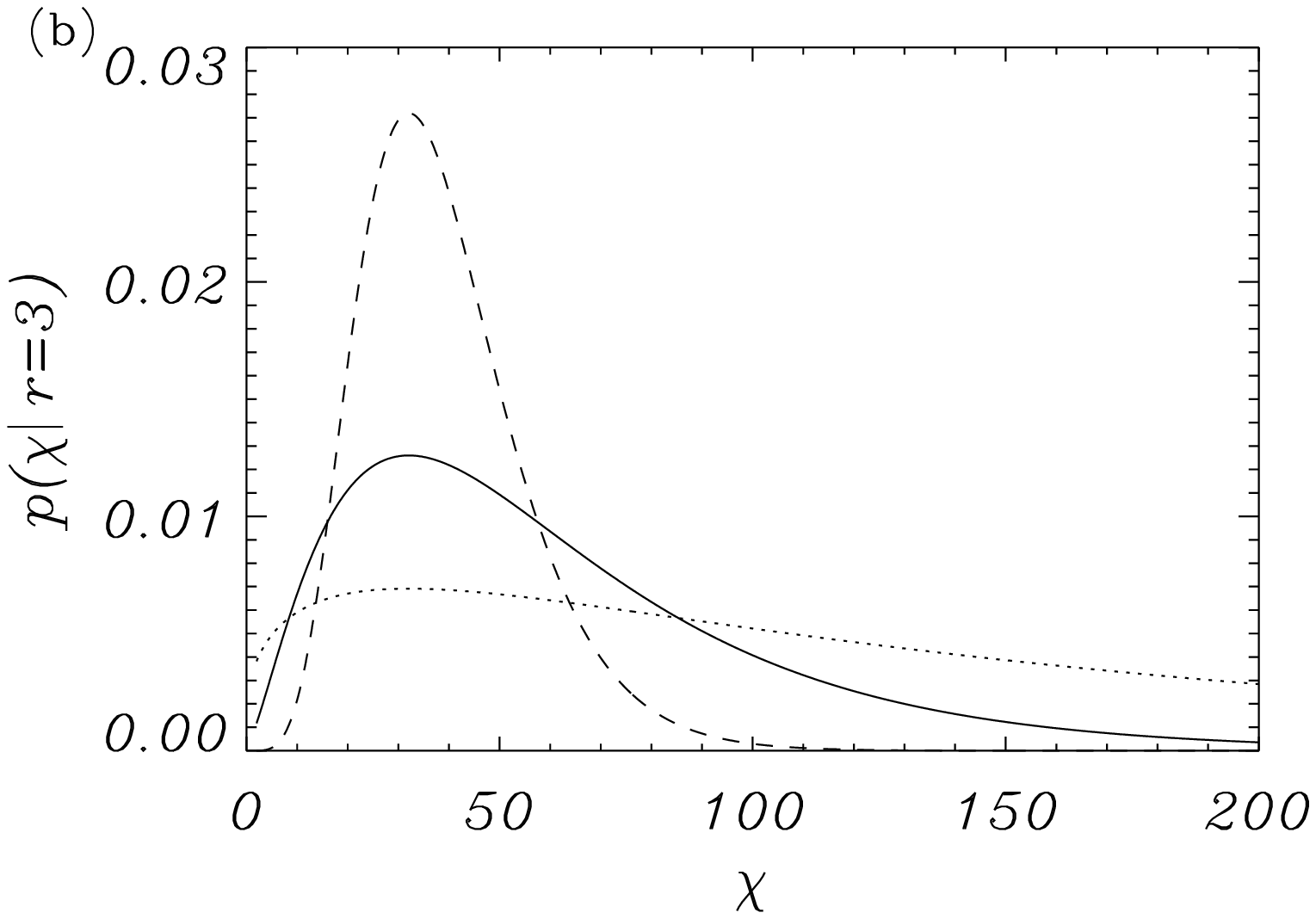}
   \caption{(a) Posterior probability distribution for $\log\chi$ under the Gaussian density gradient model, $M^G$, for 
   a period ratio of $r=3.0$ and uncertainty given by $\sigma=0.2$ (dashed); $\sigma=0.4$ (solid); $\sigma=0.8$ (dotted) and considering $\log\chi\in[1,60]$ in the prior range. For $\sigma=0.4$, the inferred median of the distribution is $\log\chi=19^{+16}_{-10}$, with uncertainty given at the 68\% credible interval. (b) Posterior probability distribution for $\chi$ under the Lorentzian density gradient model, $M^L$, for the same period ratio and uncertainty given by $\sigma=0.2$ (dashed); $\sigma=0.4$ (solid); $\sigma=0.8$ (dotted) and considering $\chi\in[2,500]$ in the prior range. For $\sigma=0.4$, the inferred median of the distribution is $\chi=51^{+47}_{-28}$.}
              \label{figinference}%
    \end{figure*}

\begin{equation}\label{rl}
r^{\rm L}=\left(\frac{P_0}{P_1}\right)^{\rm L}=1+\left(\frac{\sqrt{\chi-1}}{\arctan\sqrt{\chi-1}}\right)^{1/2},
\end{equation}

\begin{equation}\label{rg}
r^{\rm G}=\left(\frac{P_0}{P_1}\right)^{\rm G}=1+\left(\frac{2\ln\chi}{\sqrt{\pi}\ \mathrm{erf\sqrt{\ln\chi}}}\right)^{1/2},
\end{equation}
and

\begin{equation}\label{rp}
r^{\rm P}=\left(\frac{P_0}{P_1}\right)^{\rm P}=1+\left(\frac{3\chi}{2\chi+1}\right)^{1/2},
\end{equation}
where the superscripts of the period ratio indicate that the solution of the forward problem corresponds to the Lorentzian, Gaussian, and parabolic density profiles, respectively.

The three solutions for the period ratio as a function of the density variation along the thread are displayed in Figure~\ref{figforward}. When $\chi=1$, the three curves go to the period ratio $r=2$, for longitudinally homogeneous thread models \citep{soler15}. As the density variation increases the period ratio increases in the three cases. This increase is more marked for the Lorentzian profile and very subtle in the case of the parabolic profile. For the Gaussian profile an intermediate result is obtained regarding the rate of period ratio increase.

\section{Parameter inference}\label{inverse}
Using the solutions to the forward problem for the three density profiles, a method to perform parameter inference is now devised that can be used to obtain information on the density gradient along the threads from period ratio measurements. To this end, we use Bayes' theorem \citep{bayes63}, 

\begin{equation}\label{bayes}
p(\chi|r,M^i)=\frac{p(r|\chi,M^i)p(\chi|M^i)}{\int d\chi p(r|\chi,M^i)p(\chi|M^i) },
\end{equation}
to perform the inversion of parameter $\chi$ from data $r$, assuming density models $M^i$, where we choose upper-scripts
$P$, $G$, $L$ in model names M$^{P,G,L}$ as representations of the parabolic, Gaussian, and Lorentzian profiles, respectively. Bayes' theorem states that the posterior probability distribution of the unknown parameter  given the period ratio data,  $p(\chi|r,M^i)$, is proportional to the likelihood function, $p(r|\chi,M^i)$ and the prior probability, $p(\chi|M^i)$. The posterior encodes all the information about the parameter $\chi$ and quantifies the degree of belief on its values conditional on the observed data and the assumed model.

The forward solutions to the period ratio are given by Eqs.~(\ref{rl}), (\ref{rg}), and (\ref{rp}) for the Lorentzian, Gaussian, and parabolic profiles, respectively. To evaluate the likelihood for each profile, we assume the model is true. Then, the period ratio measurement ($r$) will differ from the prediction, say $r^L$ for the Lorentzian profile, because of measurement uncertainties ($e$), so that $r=r^L\pm e$. The probability of obtaining the measured value is equal to the probability of the error.  Assuming Gaussian errors, the likelihood for the three models  is then expressed in the following manner

\begin{equation}\label{like1}
p(r | \chi,M^i)=\frac{1}{\sqrt{2\pi}\sigma}\exp \left[-\frac{[r-r^i(\chi)]^2}{2\sigma^2}\right],
\end{equation}
with $i=P,G,L$ depending on the model under consideration and $\sigma^2$ the variance associated to the observed period ratio. Note that we have explicitly stated in the above expression that the likelihood is conditional on the parameter $\chi$. In the following we assign observed period ratio errors to the standard deviation $\sigma$.

The prior indicates our level of knowledge (ignorance) on the model before considering the observed data. This is translated to the level of knowledge (ignorance) on the possible values of the parameter $\chi$. We have adopted a uniform prior distribution for this parameter upon which the three model depend over a given range, so that we can write

\begin{equation}\label{prior1}
p(\chi|M^i)=\frac{1}{\chi^{\mathrm{max}}-\chi^{\mathrm{min}}} \mbox{\hspace{0.2cm}}\mbox{for}
\mbox{\hspace{0.2cm}}\chi^{\mathrm{min}}\leq \chi\leq \chi^{\mathrm{max}},
\end{equation}
and zero otherwise.  We only consider models with a positive density gradient from foot-point to tube centre, hence $\chi\geq 1$.  In the following, we adopt $\chi_{\rm min}=2$ and $\chi_{\rm max}=500$ unless otherwise stated.

The Bayesian framework enables us to perform the inference making use of all the available information in a consistent manner and with a correct propagation of errors in the observations to uncertainty on inferred parameters. Figure~\ref{figinference} shows posterior probability distributions for the density gradient along the thread computed using Eq.~(\ref{bayes}) with likelihood and prior given by Eqs.~(\ref{like1}) and (\ref{prior1}), for an hypothetical period ratio measurement, $r=3$, for three values of the period ratio measurement error and two out of the three density models, the Gaussian and Lorentzian profiles. The values of $\sigma$ have been chosen so as to be representative of period ratio measurements in which a relative error on both periods of about $10\%$ is assumed, which leads to $\Delta r=0.4$ for a value of $r=3$. For these cases, well constrained distribution are obtained that enable us to say something about the most plausible value of $\chi$. Notice the markedly asymmetric distributions in both cases, with larger upper error bars. The larger the value of $\sigma$, the less constrained are the obtained distributions. For the Gaussian profile (Fig.~\ref{figinference}a) the inversion is more conveniently done for the logarithm of $\chi$. Equation~(\ref{rg}) for this profile leads to a very slowly increasing period ratio as a function of $\chi$ (see dotted line in Figure~\ref{figforward}). For $\sigma=0.4$,  the posterior distribution has a median of $\log \chi\sim 19$, which seems unrealistically large.  For the Lorentzian profile (Fig.~\ref{figinference}b) the inversion leads to a posterior with a median of $\chi\sim51$, considering the same observational error. For the parabolic profile the inversion does not lead to a well constrained posterior distribution. The reason is that the forward solution, given by Eq.~(\ref{rp}), predicts a constant period ratio of $r=2.22$ for $\chi\gg1$. The fact that the inference for the parabolic profile fails to constrain the density gradient or that the inversion with the Gaussian profiles leads to unrealistically large values of $\chi$ does not mean that models $M^P$ and $M^G$ must be discarded, as will be next shown in the model comparison analysis. 

\begin{figure*}
\centering
   \includegraphics[width=0.49\textwidth]{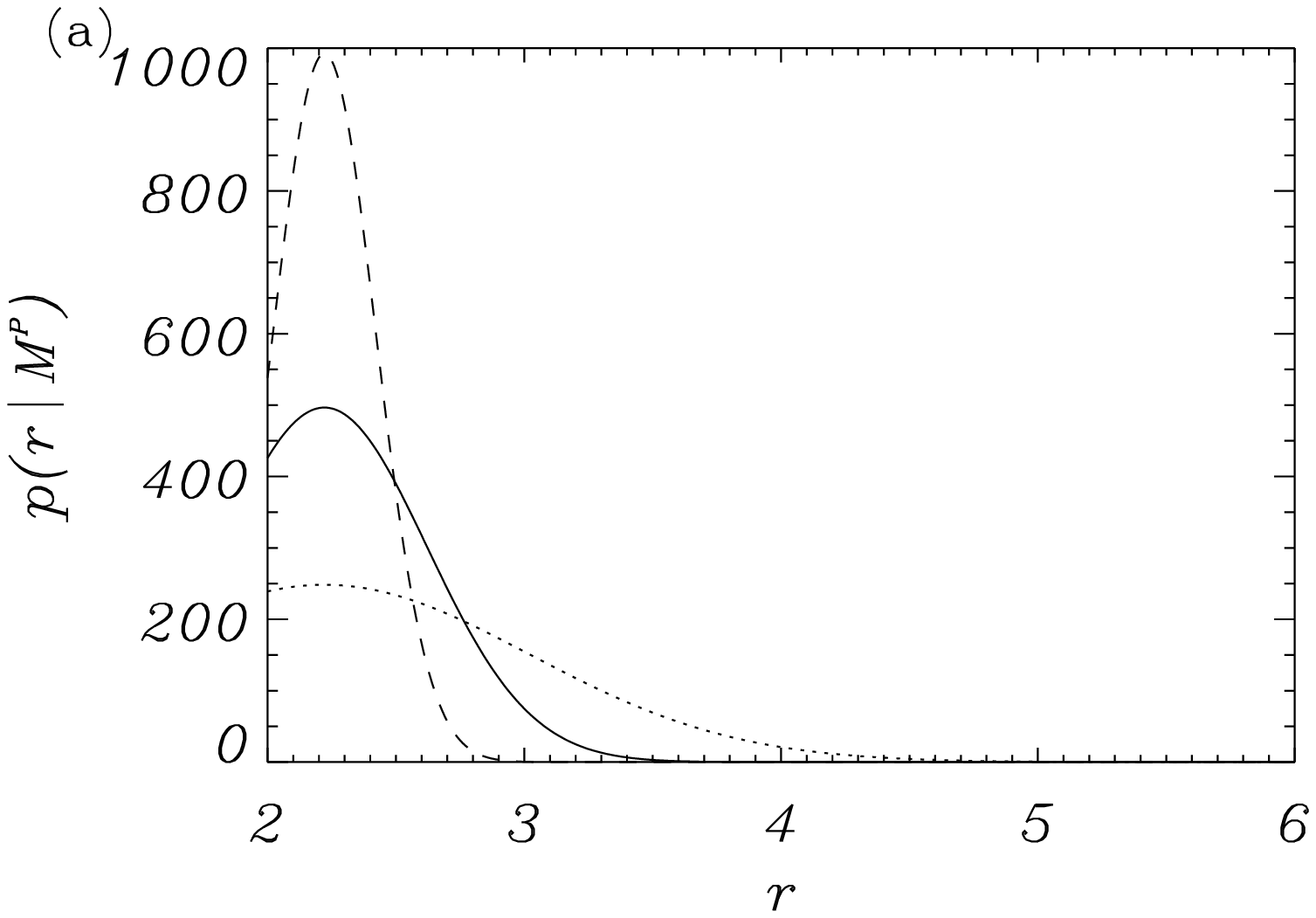}
      \includegraphics[width=0.49\textwidth]{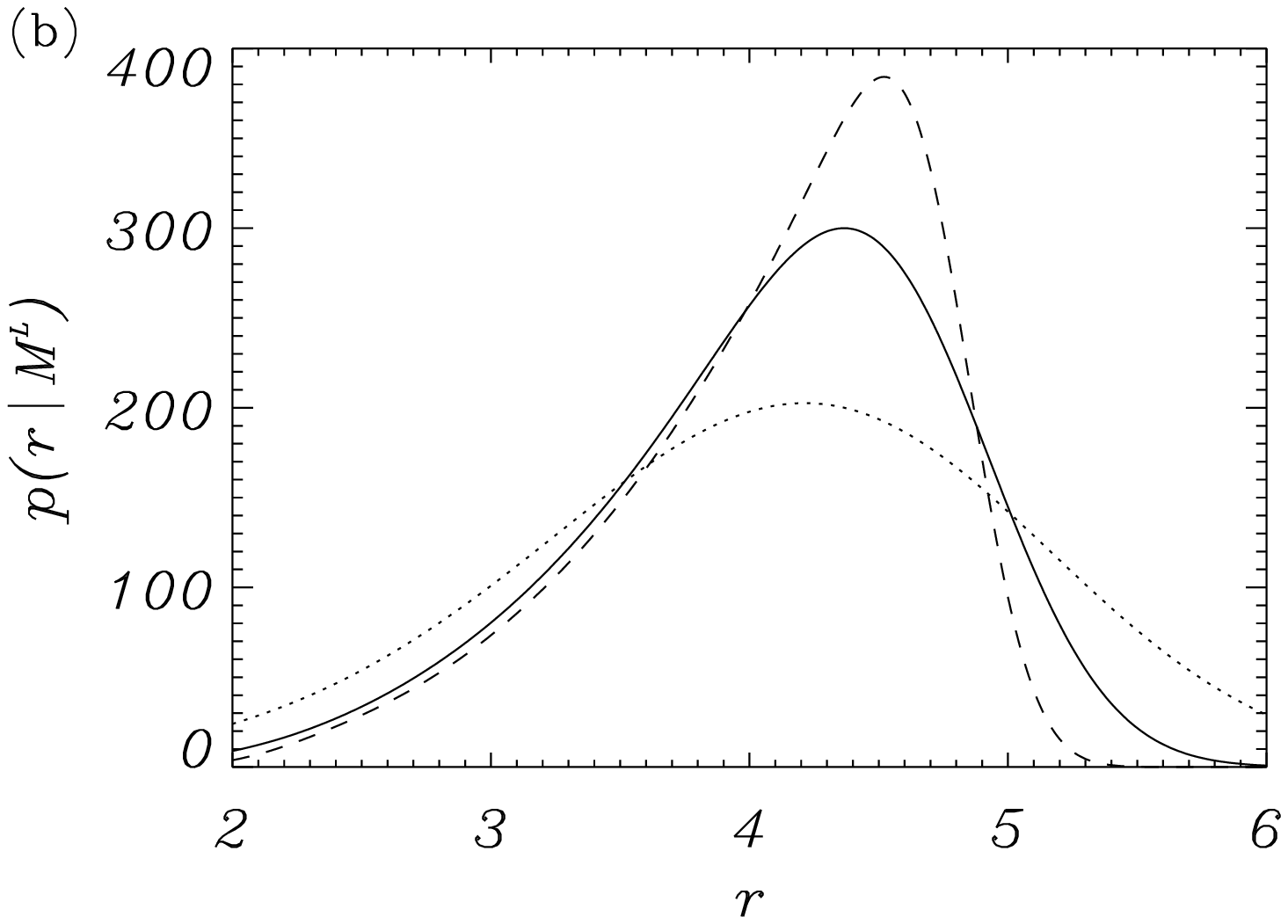}\\
   \includegraphics[width=0.49\textwidth]{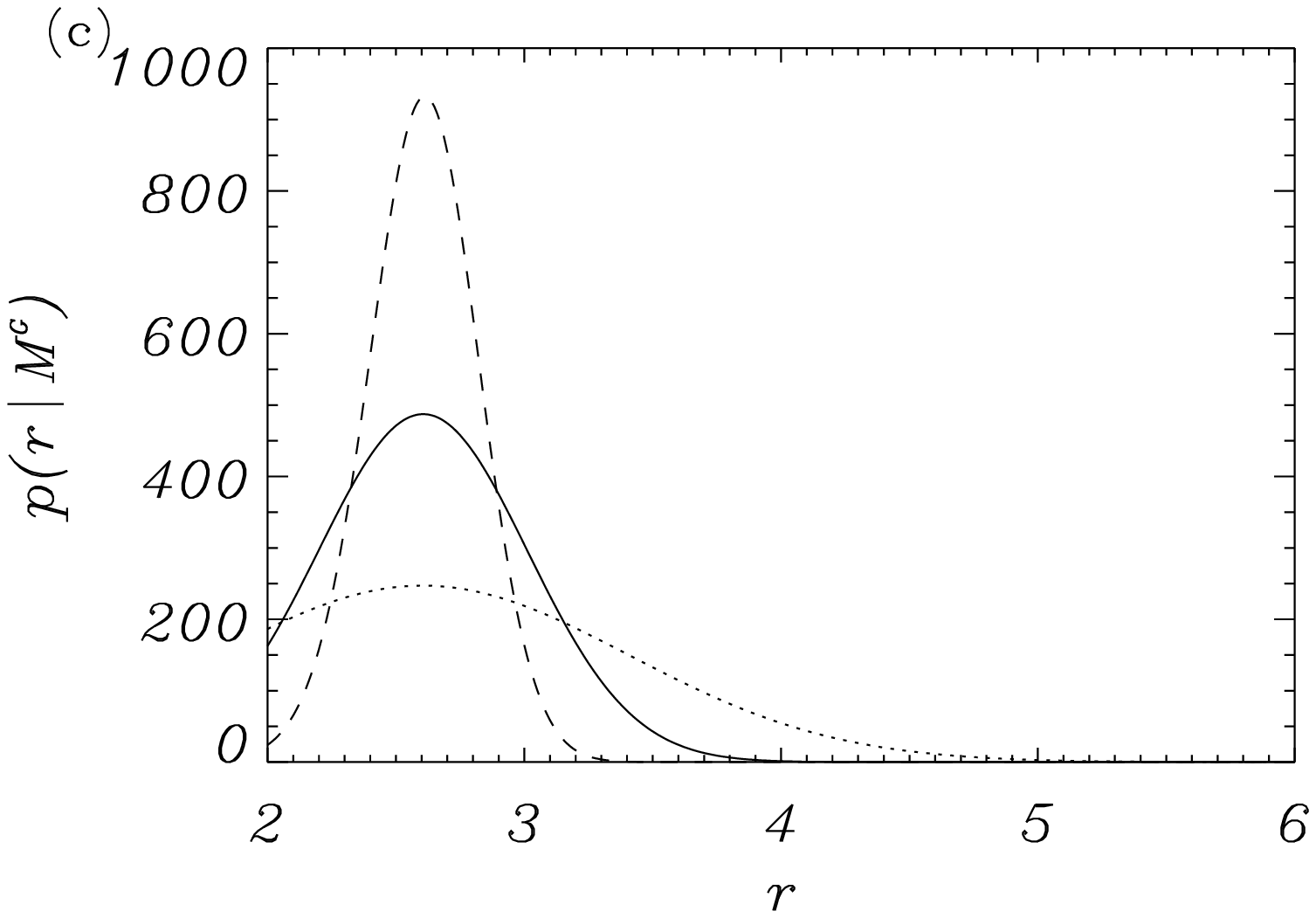}
   \includegraphics[width=0.49\textwidth]{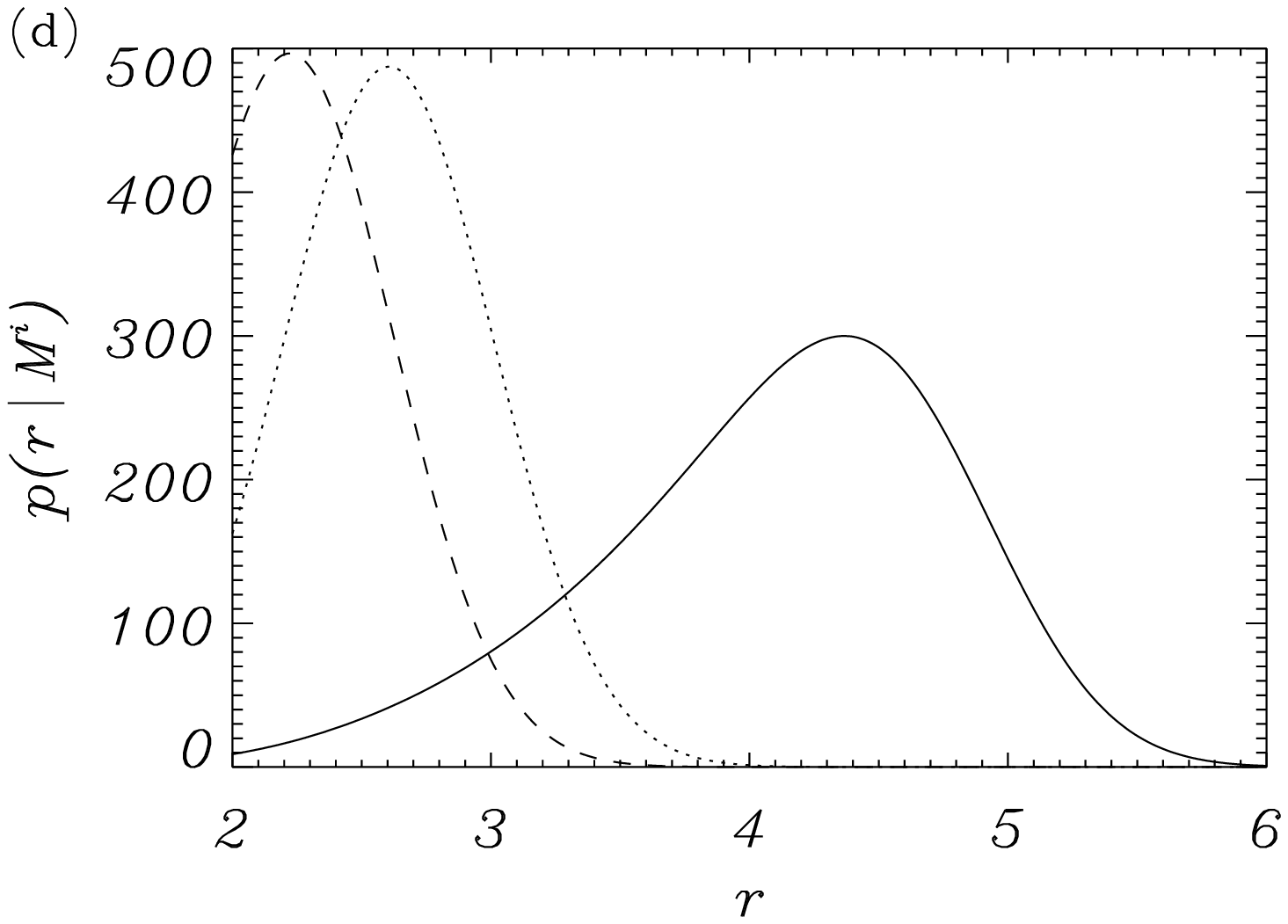}
 \caption{(a)--(c) Marginal likelihoods computed using Eq.~(\ref{marginal}) for models $M^{P}$, $M^{L}$, and $M^{G}$ as a function of the period ratio for different values of the uncertainty, with $\sigma=0.2$ (dashed); $\sigma=0.4$ (solid), and $\sigma=0.8$ (dotted). (d) The three marginal likelihoods computed for a fixed value of $\sigma=0.4$: $p(r|M^{P})$ in dashed-line; $p(r|M^{G})$ in dotted-line, and $p(r|M^{L})$ in solid-line. In all plots,  $\chi\in[2,500]$.}
              \label{marginals}%
    \end{figure*}

\section{Model comparison}\label{comparison}\label{comparison}
The Bayesian framework also enables us to present different models to the same data and assess in a quantitative manner which one is favoured by them. To develop the method to perform such a comparison, we follow a procedure similar to the one employed by \cite{arregui13a} for coronal loop oscillations in expanding and/or stratified loops.  Let us rewrite Bayes' theorem in terms of the probability of a given model, conditional on the observed data, as

\begin{equation}\label{bayes2}
p(M|D)\propto p(D|M)p(M).
\end{equation}
In order to  determine the relative plausibility of two competing models one needs to evaluate
their posterior probabilities to ascertain their relative merits. Consider two such models, $M^i$ 
and $M^j$.  By applying Eq.~(\ref{bayes2}) to them, the relative plausibility of one model against the other will be given by the posterior ratio \citep{jeffreys61}

\begin{equation}
\frac{p(M^i|D)}{p(M^j|D)}=\frac{p(D|M^i)}{p(D|M^j)}\frac{p(M^i)}{p(M^j)}.
\end{equation}
The first ratio on the right-hand side  expresses how well the observed data are predicted by model $M^i$, compared to model $M^j$. The second ratio, the prior odds ratio, measures how much our initial beliefs favoured $M^i$ over $M^j$, before considering the data. As we have no particular a priori preference for any of the three models compared in this study, we  will consider  $p(M^i)=p(M^j)=1/2$ in all our model comparisons. Our assessment of the plausibility of models will then be based on the computation of the so-called Bayes Factor of $M^i$ against $M^j$ given by

\begin{equation}\label{bf}
BF^{ij}=\frac{p(r|M^i)}{p(r|M^j)}.
\end{equation}
In our study, Bayes Factors are computed in order to assess which model, among the three proposed density profiles, better explains a given period ratio measurement. 

Before computing Bayes Factors, relevant  information concerning the plausibility of each model is obtained by first  computing their marginal likelihood, which inform us on how well a given model is in explaining data on period ratios. This is so because, in model comparison, we are interested in the most probable model, independently of the parameters, i.e., we should marginalise out all parameters. This is achieved by performing an integral of the likelihood over the full parameter space.  In our particular one-dimensional problem, the marginal likelihood for a given model $M^i$ can be written as

\begin{equation}\label{marginal}
p(r|M^i)=\int^{\chi^{\mathrm max}}_{\chi^{\mathrm min}}p(r,\chi|M^i)d\chi=\int^{\chi^{\mathrm max}}_{\chi^{\mathrm min}}p(r|\chi,M^i)p(\chi|M^i)d\chi,
\end{equation}
where $\chi\in[\chi^{\mathrm min},\chi^{\mathrm max}]$ represents the range in the parameter of the models and we have used the product rule to expand the probability of $r$ and $\chi$, given model $M^i$. 
Figure~\ref{marginals} displays marginal likelihoods for the three models under study. They are computed using expressions~(\ref{marginal}) and using the forward models (\ref{rl}), (\ref{rg}), and (\ref{rp}) for period ratios in the range 2 to 6.
In Figs.~\ref{marginals}a-c, three different values of $\sigma$ are considered to see the influence of different measurement 
error on the distribution of the plausibility for each model. Figure~\ref{marginals}d compares the marginal likelihood for the three models for a fixed period ratio measurement error. The magnitude of each of the curves provides us with the plausibility of each model, for a given observed period ratio. 

From Figs.~\ref{marginals}a-c, it is easy to see that by increasing the value of $\sigma$, related to the uncertainty on the measured period ratio, the magnitude of the three marginal likelihoods decreases and the distributions spread out over a larger range of values of $r$, thus decreasing the amount of evidence for or against any of the considered models. When comparing the marginal likelihood for the three models in Fig.~\ref{marginals}d, we see that the parabolic and Gaussian profiles are likely to produce period ratios in the lower half of the considered period ratios range,  from 2 to 4. Beyond that, their likelihood decreases significantly. According to Eq.~(\ref{rp}), see also Figure~\ref{figforward}, the parabolic profile predicts period ratios that approach the value $r=2.2$ and then keep this constant value for increasing values 
of $\chi$. The integrated marginal likelihood, dashed line in Figure~\ref{marginals}d, peaks at that value but could also explain a bit lower and larger period ratios, hence spreads both sides of the peak because of the considered uncertainty on the measured period ratio. The Gaussian profile produces a similar marginal likelihood, shifted towards larger values of the period ratio. This means that the Gaussian profile is likely to reproduce period ratio values similar to the ones reproduced by the parabolic profile. Hence, if our observed period ratio is en in range between 2 and 4, it can be difficult to obtain significant evidence for one model to be preferred over the other.  The Lorentzian profile predicts period ratios as given by Eq.~(\ref{rl}). The corresponding marginal likelihood distribution shows that this model is more likely to reproduce values of the period ratio larger than those that can be reproduced by the parabolic and the Gaussian profiles. The distribution peaks at about 4.4, but is rather extended and covers almost all values of the considered range for the period ratio.  

In order to make statements on the relative plausibility of one model over another, based on quantitative calculations, Bayes Factors are computed using Eq.~(\ref{bf}). They are functions of the marginal likelihood ratios and provide us with quantitative information on the magnitude of the evidence for one model to be preferred over another. In order to assign different levels of evidence to the values of the Bayes Factors, we use the empirical table by \cite{kass95} which assigns evidence for model $M^i$ against model $M^j$ that is {\em minimal evidence}  to values of $2\log(BF^{ij})$ in between 0 and 2; {\em positive evidence} to values in between 2 and 6; {\em strong evidence} to values in between 6 and 10; and {\em very strong evidence} to values of $2\log(BF^{ij})$ larger than 10. 

\begin{figure}
\centering
   \includegraphics[width=0.5\textwidth]{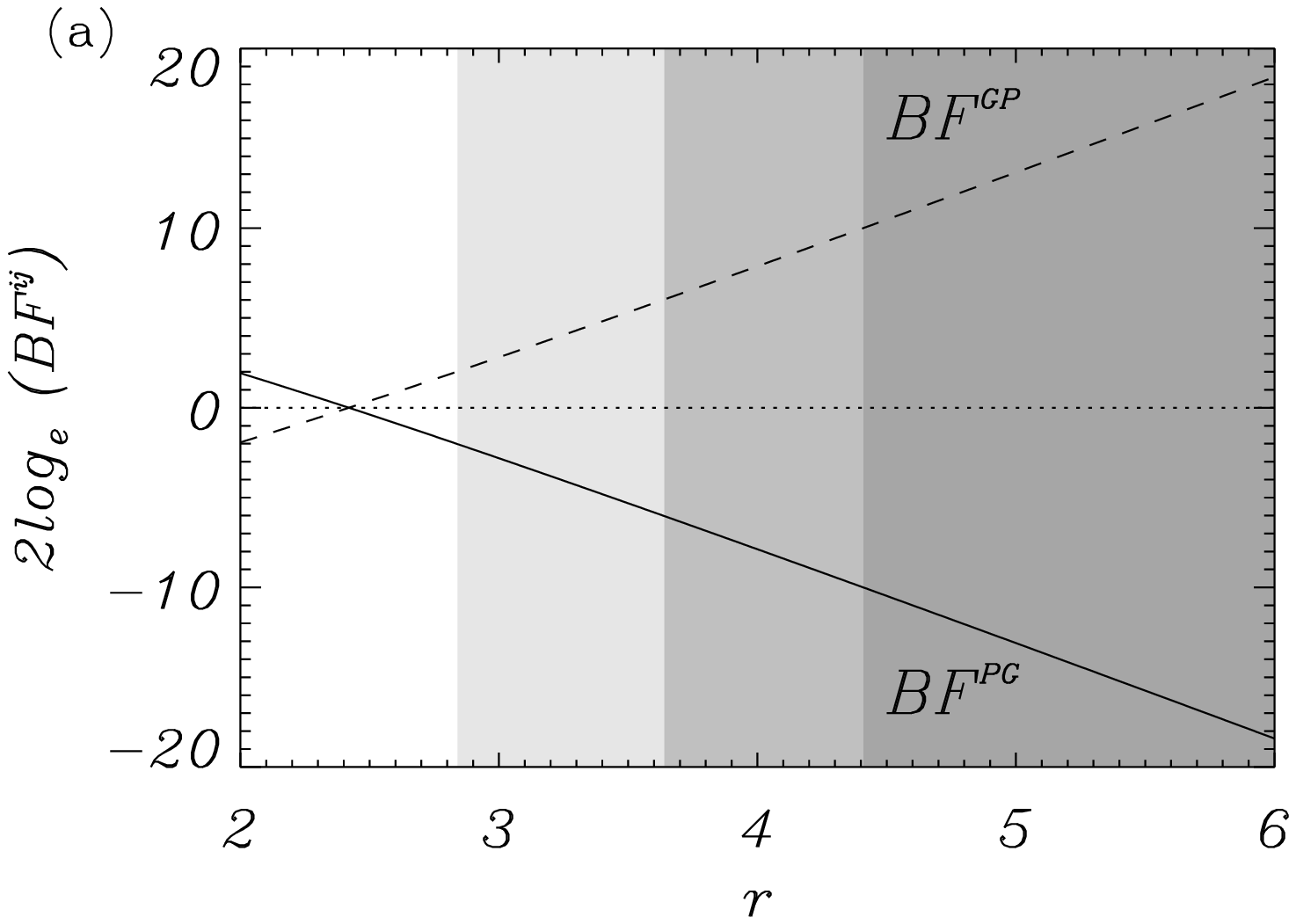}\\
      \includegraphics[width=0.5\textwidth]{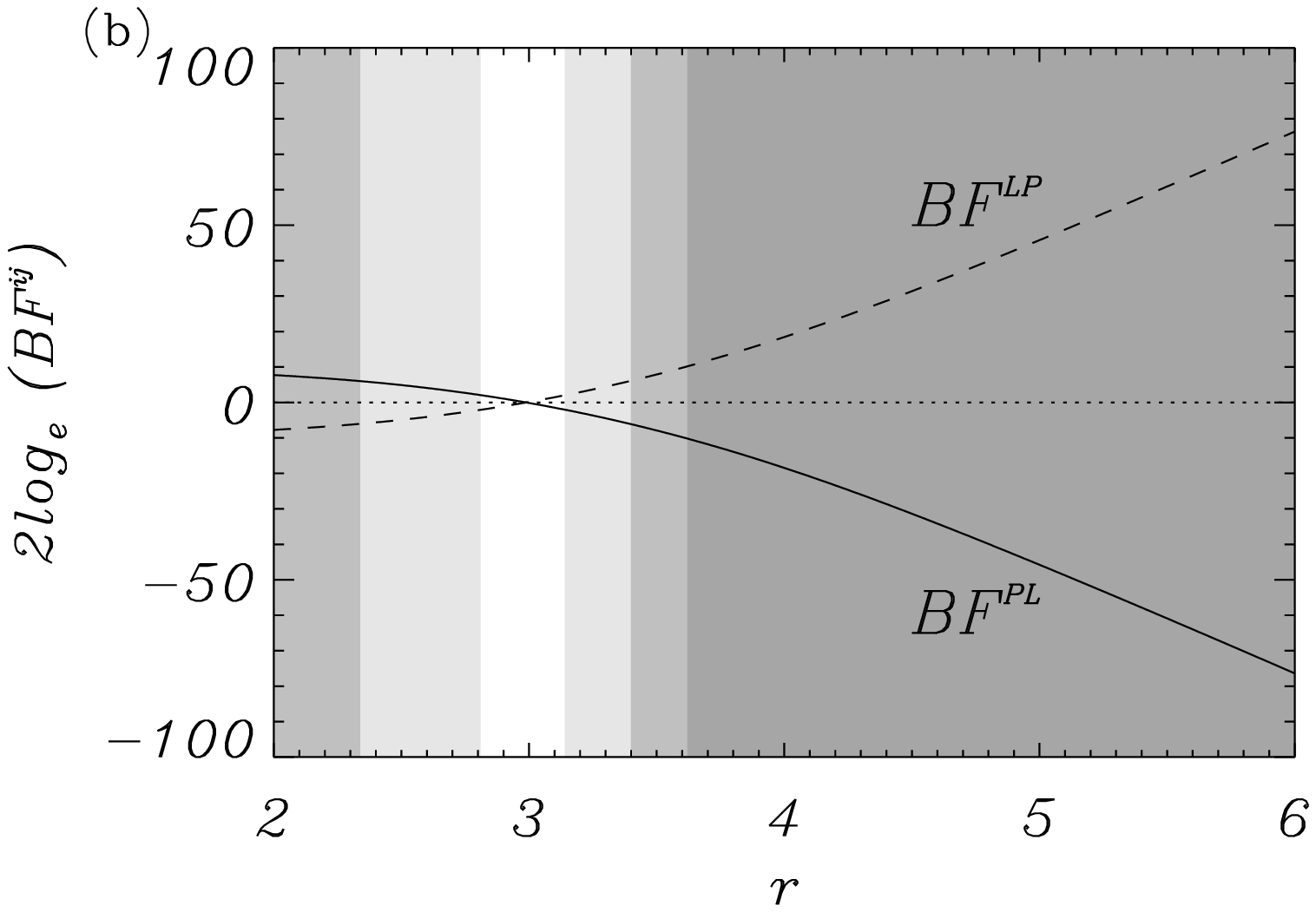}\\
   \includegraphics[width=0.5\textwidth]{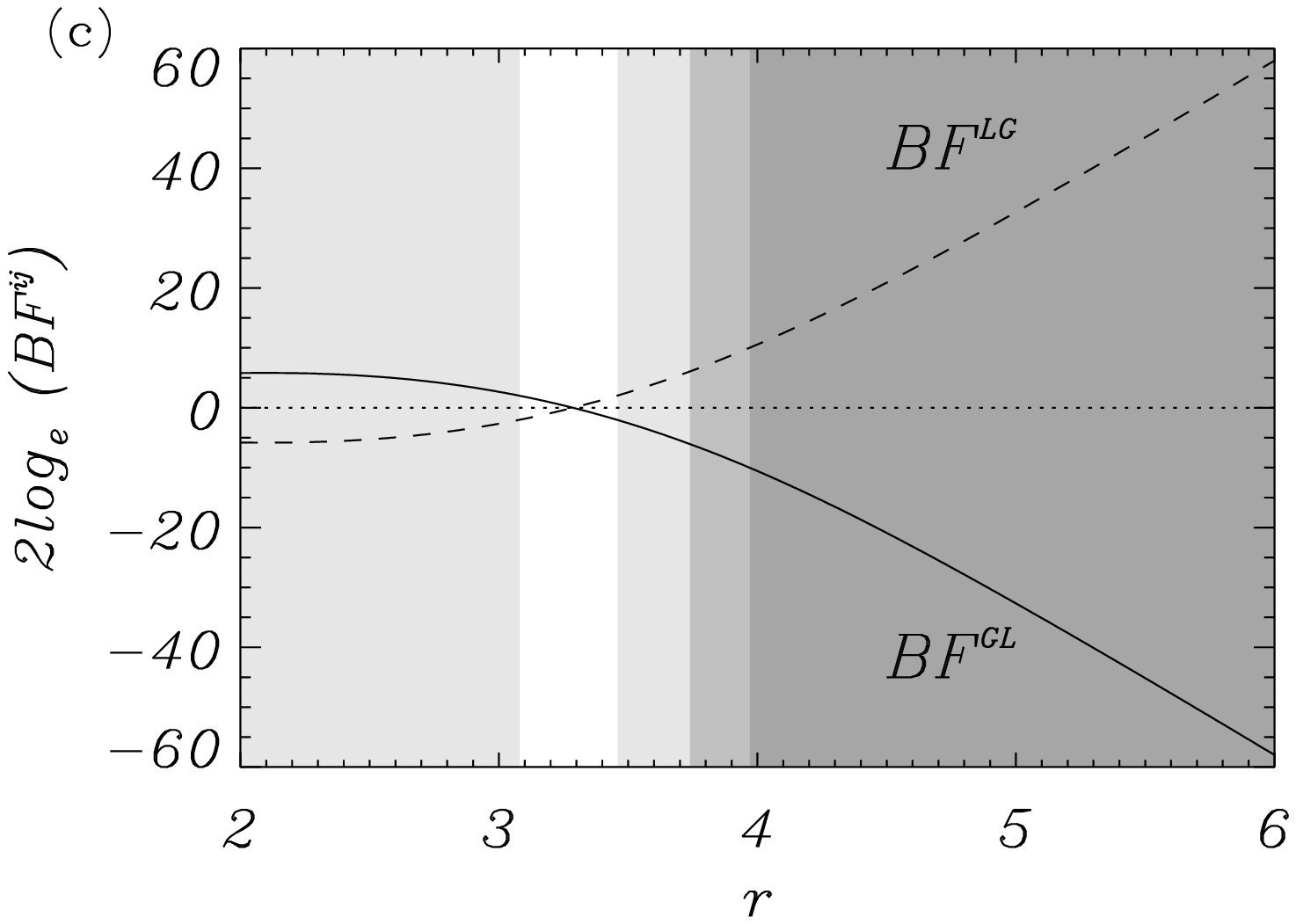}\\
 \caption{Bayes Factors computed using Eq.~(\ref{bf}) for the model comparisons between (a) the parabolic, $M^P$, and the Gaussian, $M^G$, density models; (b)  the parabolic, $M^P$, and the Lorentzian, $M^L$, density models; and (c) the Gaussian, $M^G$, and the Lorentzian, $M^L$, density models, as a function of the period ratio. A value of $\sigma=0.4$ has been considered and $\chi\in[2,500]$ in all computations.}
              \label{figbf}%
    \end{figure}

\subsection{Parabolic vs Gaussian profile}
Figure~\ref{figbf}a shows the Bayes Factors corresponding to the comparison between the parabolic and the Gaussian density models ($BF^{PG}$ in solid line and $BF^{GP}$ in dashed line), as a function of the observable period ratio.
Different shades of grey limit ranges in period ratio with different levels of evidence for one model being preferred against the alternative, depending on the magnitude of the Bayes Factors and following the scale described above. White regions indicate values of period ratio for which there is {\em minimal evidence} for any of the models that are being compared. Different shades of grey indicate regions with {\em positive}, {\em strong}, and {\em very strong} evidence, with the level of evidence being larger for darker regions.

The marginal likelihoods for the two models (see Figure~\ref{marginals}d) tell us that the Gaussian and parabolic profiles have a significant likelihood for reproducing period ratio values on a similar range, with the likelihood for the parabolic profile being slightly shifted towards smaller values of the period ratio. Bayes Factors quantify and locate exactly where and in which amount the plausibility for one model is larger than the alternative. Looking first at regions where $BF^{PG}$ is positive, meaning that the evidence for the parabolic profile is larger than that for the Gaussian profile, we find that the evidence for $M^P$ instead of $M^G$ is {\em minimal}, since $BF^{PG}$ never reaches values above 2.  The region where both Bayes Factors are too small to provide {\em positive evidence} for any of the models corresponds to the values $r=[2.0-2.84]$ (white region). Looking then at the regions where $BF^{GP}$ is positive, meaning that the evidence for the Gaussian profile is larger than that for the parabolic profile, we find {\em positive evidence} for $M^G$ instead of $M^P$ for period ratios in the range $r=[2.84-3.64]$; {\em strong evidence} in the range $r=[3.64-4.41]$; and {\em very strong evidence} for $r>4.41$.

In summary, the evidence in favour of $M^P$ instead of $M^G$ or vice-versa is {\em minimal} until we reach a period ratio of $r=2.84$. Period ratios larger than this value imply {\em positive, strong, and very strong} support for the Gaussian profile instead of the parabolic profile for which no supporting evidence can be found in the considered period ratio range.

\subsection{Parabolic vs Lorentzian profile}
Figure~\ref{figbf}b shows the Bayes Factors corresponding to the comparison between the parabolic and the Lorentzian density models ($BF^{PL}$ in solid line and $BF^{LP}$ in dashed line), as a function of the observable period ratio.
As before, the different shades of grey limit ranges in period ratio with different levels of evidence for one model being preferred against the alternative.

The marginal likelihoods for the two models (see Figure~\ref{marginals}d) tell us that the regions on measured period ratio for which each of the two models better predicts the observable are quite differentiated, with the marginal likelihood for the parabolic profile concentrated in between 2 and 4 and the marginal likelihood for the Lorentzian profile being significant for larger values of the period ratio. To what extend observed period ratios would support one model against the alternative is assessed by the magnitude of the Bayes Factors. Looking first at regions where $BF^{PL}$ is positive, meaning that the evidence for the parabolic profile is larger than that for the Lorentzian profile, we find {\em strong evidence} for $M^P$ instead of $M^L$ for period ratios in the range $r=[2.0-2.34]$ and {\em positive evidence} in the range $r=[2.34-2.81]$. Then, a region appears in white, with $r=[2.81-3.14]$, where the evidence in favour of any of the two models is 
{\em minimal} , because of the low values of both Bayes Factors. Looking then at the regions where $BF^{LP}$ is positive, meaning that the evidence for the Lorentzian profile is larger than that for the parabolic profile, we find {\em positive evidence} for $M^L$ instead of $M^P$ for period ratios in the range $r=[3.14-3.40]$; {\em strong evidence} in the range $r=[3.40-3.62]$; and {\em very strong evidence} for $r>3.62$.

The evidence for data supporting any of the competing models is inconclusive in a small range of period ratio values around $r\sim 3$, where both marginal likelihoods have similar magnitude. A measured period ratio with a value to the left of that region would support the parabolic profile, the more strongly the lower the value of the measured period ratio. A measured period ratio with a value to the right of that region would support the Lorentzian profile, the more strongly the larger the value of the measured period ratio.

\subsection{Gaussian vs Lorentzian profile}
Finally, Figure~\ref{figbf}c shows the Bayes Factors corresponding to the comparison between the Gaussian and the Lorentzian density models ($BF^{GL}$ in solid line and $BF^{LG}$ in dashed line), as a function of the observable period ratio.

The marginal likelihoods for the two models (see Figure~\ref{marginals}d) indicate that, as in previous comparison, 
the regions on measured period ratio for which each of the two models better predicts the observable are well differentiated. Looking at the values of the Bayes Factors, and first at regions where $BF^{GL}$ is positive, meaning that the evidence for the Gaussian profile is larger than that for the Lorentzian profile, we find {\em positive evidence} for $M^G$  instead of $M^L$ for period ratios in the range $r=[2.0-3.08]$. The evidence is {\em minimal} in the region $r=[3.08-3.46]$, because of the low values of the Bayes Factors. Looking then at the regions where $BF^{LG}$ is positive, meaning that the evidence for the Lorentzian profile is larger than that for the Gaussian profile, we find {\em positive evidence} for $M^L$ instead of $M^G$ for period ratios in the range $r=[3.46-3.74]$; {\em strong evidence} in the range $r=[3.74-3.97]$; and {\em very strong evidence} for $r>3.97$.

Similarly to what we found with the comparison between $M^P$ and $M^L$, the evidence in favour of $M^G$ instead of $M^L$ or vice-versa is inconclusive  in a small range of period ratio values around $r\sim 3.25$, for which their marginal likelihoods have similar magnitude. A measured period ratio with a value to the left of that region would positively support the Gaussian profile. A measured period ratio with a value to the right of that region would support the Lorentzian profile, the more strongly the larger the value of the measured period ratio. 

\subsection{Discussion of results}

Bayesian model comparison enables only to compare face to face the performance of two models at a time. However, by inspection of the previous results, we can summarise them as follows. When the period ratio is roughly in between 2 and 3, the evidence for the Lorentzian profile is lower than that for the parabolic and Gaussian profiles. However, the latter two profiles show evidence that is {\em minimal}. Hence the question cannot be settled as to which model among 
$M^P$ and $M^G$ is more plausible. When the period ratio is roughly in between 3 and 4, the evidence for the Gaussian profile  and for the Lorentzian profile are both larger than that for the parabolic profile. In turn, the Lorentzian profile has larger evidence than the Gaussian for period ratios beyond $\sim 3.5$. Finally, for larger period ratios, the Gaussian profile has larger evidence than the parabolic 
profile, but the Lorentzian profile clearly shows the largest evidence in comparison to both the Gaussian and parabolic profiles.
This means that the Lorentzian profile, with the mass density occupying a narrow extension around the central part of the flux tube offers the most plausible explanation for the density profile of prominence threads along the magnetic field if measured  period ratios are larger that $\sim 3.5$.

\cite{soler15} note that no reliable simultaneous measurements of the two periods are available yet in observations of prominence thread oscillations. \cite{lin07} report the presence of such a possible case, with a measured period ratio of $r=4.44$.  In Section 5 of \cite{soler15}, a seismology application is performed using this observed ratio, which leads to $\chi\sim347$ if the Lorentzian profile is used and $\chi\sim 10^{48}$ if the Gaussian profile is used. This leads \cite{soler15} to conclude that the measured period ratio would be compatible with the Lorentzian profile, since the inversion offers a more realistic value for the density gradient parameter. The Bayesian model comparison presented in this paper enables us to confirm this conclusion on the basis of the computed Bayes Factors. For instance, for $r= 4.44$, $BF^{GP}= 10.16 $ leading to {\em very strong evidence} supporting model $M^G$ instead of model $M^P$. However,  $BF^{LP}= 29.73$ and $BF^{LG}= 19.56$ leaving $M^L$ as the most plausible model, and with {\em very strong evidence}. 

It must be noted that these particular numbers are obtained assuming an uncertainty on the data with $\sigma=0.4$. The marginal likelihoods in Fig.~\ref{marginals} and the Bayes Factors in Fig.~\ref{figbf} depend on the uncertainty on the data. Measured errors corresponding to these events, which are not provided by \cite{lin07}, should be used to properly confirm this result. When we repeat the calculations for the same value of $r=4.41$, but doubling the error to $\sigma=0.8$, we obtain $BF^{GP}= 2.41$; $BF^{LP}= 7.03$ and $BF^{LG}=4.61$. We would have {\em positive evidence} for $M^G$ instead of $M^P$; {\em strong evidence} for $M^L$ instead of $M^P$; and {\em positive evidence} for $M^L$ instead of $M^G$. The conclusion would therefore be similar, with the evidence supporting the Lorentzian profile, but to a lesser extent.

\section{Conclusions}\label{conclusions}

Observed and theoretically modelled wave dynamics offer an alternative way to infer information about physical parameters in prominence fine structures. This seismology approach works by proposing a given model for which wave properties are computed and by inferring the unknown plasma and field conditions from a comparison with observed wave properties. Alternative models are usually proposed and they might lead to different inversion results.  This makes necessary to obtain information about which one among alternative models better explains observed data. 

Bayesian analysis techniques offer the only self-consistent method to compare theoretical models and observations and to propagate uncertainty from measured quantities to inferred parameters or to amount of evidence supporting a given model. We have applied Bayesian inference and model comparison techniques to obtain information on the density structure along solar prominence threads and to assess which one among three alternative models for the density variation along prominence threads would better explain given observed values for the period ratio. 

The three density models offer different results for the inversion of the density gradient parameter. A Lorentzian density profile, with plasma density concentrated around the centre of the tube seems to offer the most realistic inversion result. A Gaussian profile, with the mass spread over a larger fraction of the tube, would require unrealistically large values of the density gradient parameter. A parabolic density distribution does not enable us to obtain well constrained posterior probability distributions for the density gradient. However, this does not imply that the parabolic and Gaussian models can be disregarded. 

Our model comparison results indicate that for period ratios roughly in between 2 and 3, the parabolic and Gaussian profiles lead to the largest marginal likelihoods. By computing Bayes factors we have obtained  a full quantitative assessment on the plausibility of each model as a function of the observed parameter considering all the information available on data, their uncertainty, and the model parameters. The different levels of evidence for different ranges of the observable are obtained on the basis of the magnitude of the Bayes Factors. They depend on the specific models and the uncertainty on the data and are, therefore, by no means presented as general limits. The procedure should be repeated for every observation. Our results indicate that a Lorentzian profile, with the mass density concentrated around the centre of the magnetic flux tube, would offer the most plausible explanation for measurements in which the period ratio deviates significantly from its value of 2 in homogeneous thread models. For period ratio measurements roughly in between 2 and 3, the evidence would be inconclusive and not sufficient to decide between the Gaussian and the parabolic profiles.

The models used in this study consider static threads and  do not take into account the effect of flows. Results from \cite{soler11b} on flowing threads suggest that the period ratio may change as the dense part of the thread flows along the magnetic flux tube. 

Although no observations of the period ratio in prominence threads are currently available, the framework developed in this article has the potential to infer information about the spatial variation of density along threads if observations of the period ratio are eventually reported.

 \begin{acknowledgements}
We are grateful to the referee for comments that improved the manuscript. We acknowledge financial support from the Spanish Ministry of Economy and Competitiveness (MINECO) through projects AYA2011--22846 (Dynamics and Seismology of Solar Coronal Structures), AYA2014-55456-P (Bayesian Analysis of the Solar Corona), from FEDER funds, and from CAIB through the ``Grups Competitius'' program.  IA acknowledges financial support through a Ram\'on y Cajal fellowship. RS acknowledges support from MINECO through a Juan de la Cierva grant, from MECD through project CEF11-0012, and from the ``Vicerectorat d'Investigaci\'o i Postgrau'' of the Universitat de les Illes Balears.
\end{acknowledgements}

\end{document}